\documentstyle[prl,aps,twocolumn,epsfig]{revtex} 
\begin{document}
\preprint{} 
\draft 

\title{ Soliton in BCS superfluid Fermi gas }

\author{ Jacek Dziarmaga and Krzysztof Sacha }

\address{
Intytut Fizyki Uniwersytetu Jagiello\'nskiego, 
ul.~Reymonta 4, 30-059 Krak\'ow, Poland 
}

\date{ June 23, 2004 }

\maketitle

\begin{abstract}
We analyze a superfluid state of two species gas of
fermions trapped in a quasi-1D harmonic potential and
interacting via attractive s-wave collisions. It is shown
that the gap equation posesses a self consistent solution
with an antisymmetric gap function. The gap function has a
localized soliton or kink in the center of the trap.
\end{abstract}

\pacs{03.65.-w, 03.75.Fi, 42.50.Md} 

Over the last decade we experienced rapid development
in trapping and cooling of dilute atomic gases that
made condensates of bosonic atoms standard objects of
investigation in laboratories of atomic physics
\cite{anglin02n}. At the same time transition to a
superfluid state in cold fermionic gases seems to be
already attainable \cite{Feshbach,Timmermans,jin}.
The remarkable level of manipulation and control of
dilute bosonic gases allowed for a number of
spectacular experiments with Bose-Einstein
condensates (BEC) \cite{anglin02n}. Similar
experimental activity in fermionic superfluidity may
be expected in the immediate future.

In the present Letter we investigate superfluid phase
of a two-species fermionic gas in the weak coupling
BCS regime trapped in a quasi one-dimensional (1D) harmonic 
trap. The presence of a trapping potential is a necessary condition
to have superfluidity in an effective one dimensional
system. We show that the gap function can have a
soliton. Dark solitons in a 1D Bose-Einstein condensate
have been investigated theoretically and
experimentally for a few years \cite{anomalous,DS}.
Atomic density drops to zero in the dark soliton.
In contrast, the quasi 1D Fermi superfluid can 
support a soliton that does not show up in the
density distribution but only in the gap function. 

Two species of Fermions $\hat\psi_-(x)$ and
$\hat\psi_+(x)$ with an attractive mutual s-wave
interactions in a quasi-1D harmonic trap are
described by the Hamiltonian
\begin{eqnarray}
\hat H~=~
\int_{-\infty}^{+\infty} dx~
\left[
\hat\psi^{\dagger}_+ {\cal H}_0 \hat\psi_+ +
\hat\psi^{\dagger}_- {\cal H}_0 \hat\psi_- 
-g 
\hat\psi^{\dagger}_- \hat\psi^{\dagger}_+ 
\psi_+ \psi_-
\right]~.
\nonumber
\end{eqnarray}
Here ${\cal H}_0=-\frac12\frac{\partial^2}{\partial
x^2}+\frac12 x^2-\mu$ is the single particle Hamiltonian in the
trap units. $g>0$ is an effective 1D strength of
attraction. In the usual BCS mean-field
approximation with the pairing field
$\Delta(x)=g\langle\hat\psi_+(x)\hat\psi_-(x)\rangle$ and the
Hartree-Fock potential
$W(x)=-g\langle\hat\psi^{\dagger}_{\pm}(x)\hat\psi_{\pm}(x)\rangle$
the stationary problem at zero temperature is equivalent to
solving the Bogoliubov-de Gennes equations \cite{cegla}
\begin{eqnarray}
\omega_m u_m &=& 
+{\cal H}_0 u_m + W(x)u_m + \Delta(x)   v_m ~,\nonumber\\
\omega_m v_m &=& 
-{\cal H}_0 v_m - W(x)v_m + \Delta^*(x) u_m ~,\label{BdG}
\end{eqnarray}
together with the self-consistency conditions
\begin{eqnarray}
\Delta(x) &=& g \sum_{m=1}^{\infty} u_m(x)v_m^*(x) ~,\label{Deltadef}\\
W(x)      &=& -g \sum_{m=1}^{\infty} |v_m(x)|^2    ~.\label{Wdef}
\end{eqnarray}
Unlike in 2D or 3D \cite{UV,LDA}, in 1D both sums are convergent.  
Similar Bogoliubov-de Gennes equations with soliton solutions appear 
in the theory of conducting polymers \cite{Hegger} where the order
parameter $\Delta$ is real and lives on a dicrete 1D lattice.

In numerical calculations the infinite sums must be replaced by
finite sums up to a certain cut-off $\Lambda$,
\begin{equation}
\Delta_{\Lambda}(x) = 
g \sum_{m=1}^{\Lambda} u_m(x)v_m^*(x)~. 
\label{goleg}
\end{equation}
For a sufficiently large cut-off $\Delta_{\Lambda}(x)$
converges to $\Delta(x)$. In principle for fixed 
$(u_m,v_m)$ the difference is
$\Delta(x)-\Delta_{\Lambda}(x)={\cal O}(\Lambda^{-1/2})$.
Unfortunately, we found that the small error
${\cal O}(\Lambda^{-1/2})$ was amplified by the nonlinear
set of Eqs.(\ref{BdG})-(\ref{Wdef})
and the convergence was much slower. This is why
we use a different prescription
\begin{equation}
\Delta(x) =
\Delta_{\Lambda}(x) +
\delta\Delta_{\Lambda}(x)~. \label{DeltaLambda}
\label{niegoleg}
\end{equation}
Here $\delta\Delta_{\Lambda}(x)\approx g
\sum_{m=\Lambda+1}^{\infty}u_m(x)v_m^*(x)$ is an
approximate analytical expression for the part of the
infinite sum that is normally ignored. Again for
$\Lambda$ large enough we recover the infinite sum,
but now the convergence is much faster than for
the simple cut-off in Eq.(\ref{goleg}).

We use the local density approximation (LDA) to 
evaluate the high energy part of the gap function
\begin{equation}
\delta\Delta_{\Lambda}(x)=
g
\int_{|k|>k_{\Lambda}(x)} \frac{dk}{2\pi}~
\frac{\Delta(x)}{2E(k,x)}~.
\label{deltaLDA}
\end{equation}
In the LDA the superfluid is considered locally 
uniform so that locally one can use the BCS theory 
for a uniform system. In equation (\ref{deltaLDA}) 
the 
$E(k,x)=\sqrt{\Delta^2(x)+\epsilon^2(k,x)}$
is a local quasiparticle energy with
$\epsilon(k,x)=\frac{k^2}{2}+\frac{x^2}{2}+W(x)-\mu$.
A local cut-off momentum $k_{\Lambda}(x)$
corresponding to the cut-off energy
$\omega_{\Lambda}+\mu$ is given by
$k^2_{\Lambda}(x)=2\omega_{\Lambda}+2\mu-x^2-2W(x)$
when positive and zero otherwise. A combination of
Eqs.(\ref{DeltaLambda}) and (\ref{deltaLDA}) gives
to leading order in $\Delta(x)$
\begin{eqnarray}
\Delta(x) &\approx&
g_{\Lambda}(x) \sum_{m=1}^{\Lambda} u_m(x)v^*_m(x)~,
\label{Deltaren}\\
\frac{1}{g_{\Lambda}(x)} &\stackrel{\rm def.}{=}&
\frac{1}{g}-
\frac{1}{2\pi}\int_{k_{\Lambda}(x)}^{\infty}\frac{dk}{2E(k,x)}
\nonumber\\
& \approx &
\frac{1}{g}-
\frac{1}{2\pi k_F(x)}
\ln\left(
\frac{k_{\Lambda}(x)+k_F(x)}{k_{\Lambda}(x)-k_F(x)}
\right)~.
\nonumber
\end{eqnarray}
Here $k_F(x)=\sqrt{2\mu-x^2-2W(x)}$ is a local Fermi
momentum.  The LDA applied to the highest
quasiparticle states results in an $x$-dependent
attraction strength $g_{\Lambda}(x)$. The enhanced
$g_{\Lambda}(x)>g$ approximately compensates for the
high energy quasiparticle states missing in the
finite sum. We use the LDA to evaluate only the sum
$\delta\Delta_{\Lambda}(x)$ over the high energy
states. Although accuracy of LDA as such may be
disputable, there is little doubt that it is a very
accurate approximation in the high energy regime
where $k_{\Lambda}(x)$ is much greater than any rate
of variation of $\Delta(x)$.

\begin{figure}
\vspace*{0.5cm}
\epsfig{file=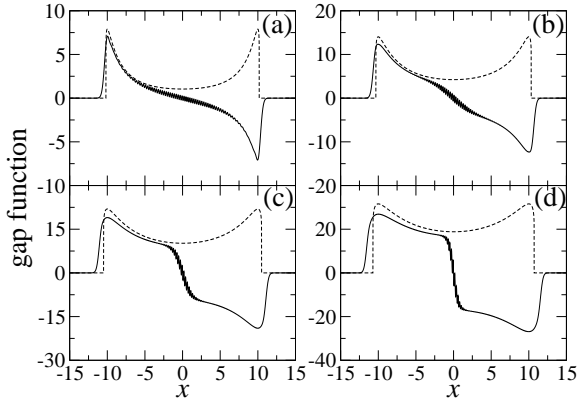, width=8.6cm, clip=0.5cm}
\caption{ 
Gap functions $\Delta(x)$ in the antisymmetric state
(solid lines) and approximate LDA gap functions
$\Delta_{\rm LDA}(x)$ in the symmetric ground state
(dashed lines) for $g=6$ [panel (a)], $g=8$ [panel
(b)], $g=10$ [panel (c)], and $g=12$ [panel (d)]. The
chemical potential is $\mu=50$ which gives an average
total number of atoms ranging from $2\times 70$ for
$g=6$ to $2\times 82$ for $g=12$.
}
\label{FigDeltas}
\end{figure}

In this paper we consider an antisymmetric
$\Delta(x)=-\Delta(-x)$ with a symmetric density of
atomic cloud $W(x)=W(-x)$ in the weak coupling BCS
regime. The solid lines in Fig.\ref{FigDeltas} show
the antisymmetric gap functions for four values of
the attraction strength $g$. Panels (b)-(d) show very
clear characteristic soliton (kink) patterns, but in
panel (a) the width of the soliton is comparable to
the size of the atomic cloud and we see just a plain
antisymmetric state.

The dashed lines in Fig.\ref{FigDeltas} show symmetric
ground state $\Delta(x)$'s obtained with the LDA. To get
this approximate gap function we set $\Lambda=0$ on the
left hand side of Eq.(\ref{deltaLDA}) which replaces the
$\delta\Delta_{\Lambda}(x)$ by an approximate $\Delta(x)$.
Then we solve Eq.(\ref{deltaLDA}) to leading order in
$\Delta(x)$:
\begin{equation}
\Delta_{\rm LDA}(x)~=~
4~k_F^2(x)~
e^{-\pi k_F(x)/g}~
\label{DeltaLDA}
\end{equation}
wherever $k_F(x)$ is real and $0$ otherwise. The
agreement between the symmetric $\Delta_{\rm LDA}$
and modulus of antisymmetric gap functions is
reasonable even for the small numbers of atoms
considered in Fig.\ref{FigDeltas}. The solitons in
panels (b)-(d) can be considered as localized defects
in a locally uniform superfluid.

\begin{figure}
\epsfig{file=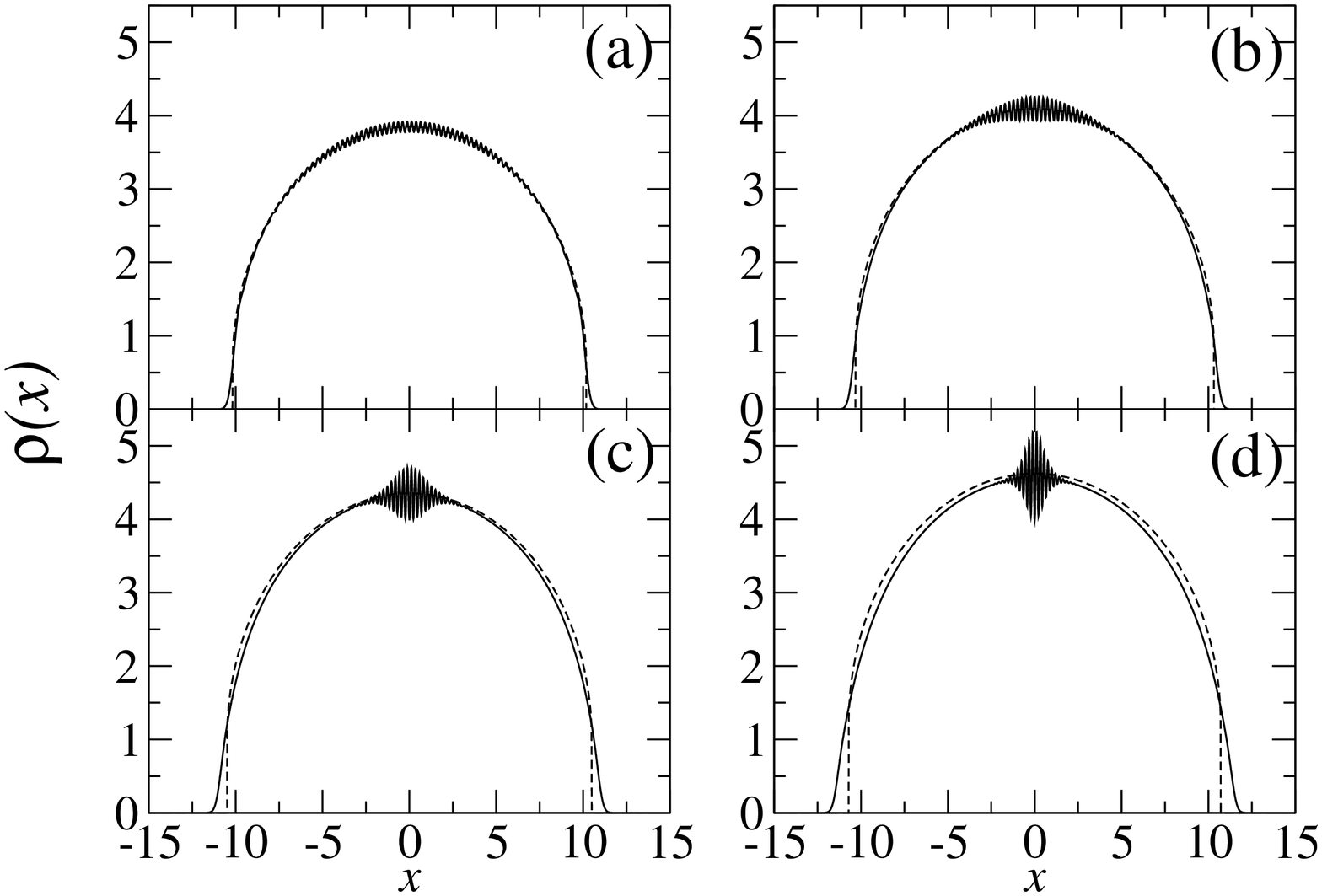, width=8.6cm, clip=}
\vspace*{0.8cm}
\caption{
Atomic densities $\rho(x)=
\langle\hat\psi^{\dagger}_{\pm}(x)\hat\psi_{\pm}(x)\rangle$ corresponding to the
antisymmetric states in Fig.\ref{FigDeltas}.
The fast oscillations of $\rho(x)$ in the
solitons average to zero on length scales longer
than $k_F^{-1}$. The dashed plots show approximate
$\rho_{\rm LDA}=\frac{g}{\pi^2}+
\frac{1}{\pi}\sqrt{2\mu-x^2+\frac{g^2}{\pi^2}}$.
}
\label{FigRhos}
\end{figure}

The gap function $\Delta(x)$ has a soliton but,
unlike for dark solitons in BEC \cite{anomalous,DS},
corresponding atomic density shows hardly any
symptoms of the soliton at all, see
Fig.\ref{FigRhos}. The soliton in the weakly
interacting BCS superfluid is not a dark soliton.

\begin{figure}
\epsfig{file=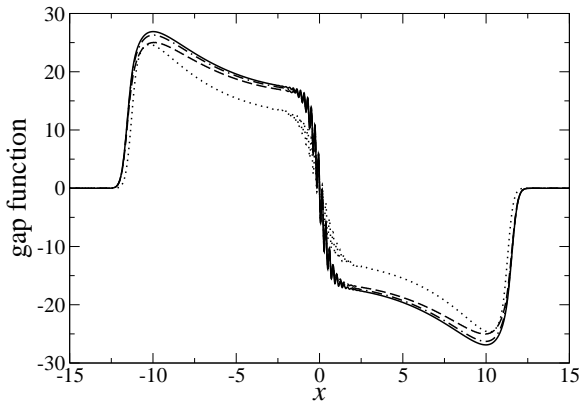, width=8.6cm, clip=}
\caption{
Gap functions calculated in different approximations for chemical
potential $\mu=50$ and attraction strength $g=12$. Dotted line corresponds to 
Eq.~(\ref{goleg}) with $\Lambda=1000$. Dashed, dotted-dashed and solid lines
are the results that base on Eq.~(\ref{Deltaren}) with $\Lambda$ equal to 400, 
700 and 1000, respectively.}
\label{FigLambda}
\end{figure}

To demonstrate convergence of the scheme based on
Eq.~(\ref{Deltaren}) we show $\Delta(x)$ for $g=12$
and increasing value of $\Lambda$ in
Fig.\ref{FigLambda}. In this figure we also show
results corresponding to Eq.~(\ref{goleg}) where the
contributions from the high energy states are
neglected. We see that the results based on the
scheme Eq.~(\ref{Deltaren}) are closer to the true
solution, even for $\Lambda=400$, than those based on
the simple cut-off scheme (\ref{goleg}) with
$\Lambda=1000$.

\begin{figure}
\epsfig{file=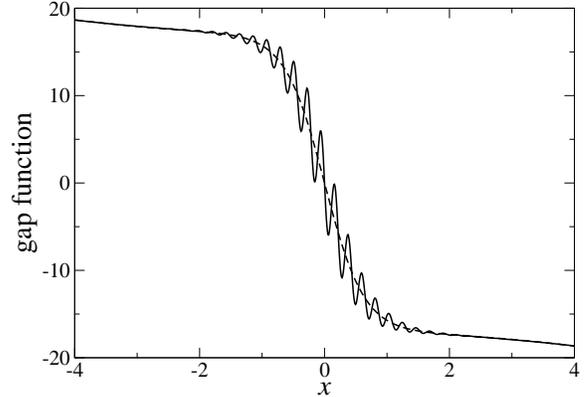, width=8.6cm, clip=}
\caption{
Solid line indicates the gap function shown in panel (d) of 
Fig.~\ref{FigDeltas}. Dashed line is a fit to the solid line with a trial
function $\gamma\Delta_{\rm LDA}(x)\tanh(x/\xi)$ perfomed in 
the range of space presented in the figure only. 
The best fit was obtained for $\alpha=1.07$, Eq.~(\ref{alp}).
}
\label{FigFit}
\end{figure}

We need some analytical insights to extrapolate the
numerical results to a large number o atoms $N$. The
most intriguing issue is the width of the soliton. A
simple estimate of the healing length $\xi$ can be
obtained in the LDA. The local quasiparticle energy
$E(k,x)$ can be expanded in small fluctuations
$\delta k=k-k_F$ around Fermi momentum
\begin{equation}
E=
\sqrt{\Delta^2+\left(\frac{k^2-k_F^2}{2}\right)^2}
\approx
k_F\sqrt{\frac{\Delta^2}{k_F^2}+\delta k^2}~.
\end{equation} 
We see that the quasiparticle spectrum is affected by the
gap only in the range of $\delta k\approx \Delta/k_F$.
This is the dispersion of momenta available to construct a
localized soliton and so the healing length must scale as
\begin{equation} 
\xi~=~\alpha~\frac{\sqrt{2\mu}}{\Delta}
\label{alp}
\end{equation} 
with a constant $\alpha={\cal O}(1)$. To estimate the
constant we fit the solitons in panels (b)-(d) in
Fig.\ref{FigDeltas} with a trial function
$\gamma\Delta_{\rm LDA}(x)\tanh(x/\xi)$, where $\gamma$
and $\xi$ are fitting parameters.  The best fit to
the central part of panel (d) is shown in
Fig.\ref{FigFit}. From $\xi$'s we obtain the best
fits for $\alpha$'s and show them in Table~\ref{t1}
--- it is fair to say that $\alpha\approx 1$.

Another interesting question is when the soliton is a
nice localized kink as in panels (b)-(d) of
Fig.\ref{FigDeltas} and when its width is comparable
or exceeds the width of the atomic cloud as in panel
(a). A simple estimate of the ``kinkness'' of the
antisymmetric state is the ratio of the healing
length to the Thomas-Fermi width of the cloud $x_{\rm
TF}\approx\sqrt{2\mu}$,
\begin{equation}
\frac{\xi}{x_{\rm TF}}\approx
\frac{1}{\Delta}~.
\end{equation}
This factor can be made small for large $N$ when a
large $\Delta\gg 1$ is easily compatible with the
weak coupling condition $\Delta\ll\mu$ necessary for
the LDA to be accurate.

Once we have a localized soliton another interesting
question arises. A localized soliton should behave
like a non-dispersive point particle. As the atomic
cloud in the trap is non-uniform this particle feels
an effective potential $V_{\rm eff}(q)$ with soliton
position $q$. The question is what is the shape of
this potential. This is an important issue related to
stability because, for example, the dark solitons in
BEC experience an inverted harmonic potential
\cite{anomalous}. Their small fluctuation spectrum
has an anomalous mode with negative frequency which
might imply their instability. Fortunately, it does
not for technical reasons specific to the dark
solitons in harmonic traps. To get an insight into 
$V_{\rm eff}(q)$ we make a simple estimate of the soliton
energy. According to Ref.\cite{cegla} the difference
between the energy density of the superfluid state
with a gap $\Delta\neq 0$ and the normal state with
$\Delta=0$ is given by
\begin{equation}
\delta\varepsilon=
\int_0^{\Delta} d\Delta^{'}~
\left( \Delta^{'} \right)^2~
\frac{d~g^{-1}}{d\Delta^{'}}~.
\end{equation}
When combined with the LDA dependence of the gap
on the attraction strength $g$ in 
Eq.(\ref{DeltaLDA}), this formula gives 
$\delta\varepsilon=-\Delta^2/2\pi k_F$. Now, in zero
order approximation the soliton is an area of width
$\xi$ where this energy gain is missing, hence the
energy of the soliton is roughly
$-\delta\varepsilon\times\xi$ or 
\begin{equation}
V_{\rm eff}(q)~\sim~
\left|\Delta_{\rm LDA}(x=q)\right|~. 
\end{equation}
We come to a remarkably simple conclusion that the
effective potential is proportional to the dashed
plots in Fig.\ref{FigDeltas}. Small fluctuations of
the soliton close to the minimum of the potential in
the center of the trap have positive frequency. There
is no anomalous mode that might imply thermodynamic
instability. In order to deexcite the superfluid from
the soliton state to the symmetric ground state the
soliton has to be pushed beyond the potential barrier
at the edge of the atomic cloud. In this sense
solitons in the BCS Fermi superfluid are more robust
than dark solitons in BEC. Hopefully the dark soliton
experiments in BEC will find their counterpart in
BCS Fermi superfluid.

As was shown in Ref.\cite{Baranov}, in a 3D Fermi
superfluid in an isotropic harmonic trap there is no
energy gap in the quasiparticle energy spectrum even
for a large gap function $\Delta(\vec x)$ in the
center of the trap. 
Thus in 3D the
spectroscopic detection of the BCS state requires the
probing laser beam to be focused on the locally
uniform cenral part of the trap. However, the
detection schemes based on the measurement of the
energy gap \cite{detection} can be very useful in 1D.
In 1D an nonzero $\Delta(x)$ directly translates into
a gap in the quasiparticle spectrum. 

What is more the same schemes can be used to detect
the soliton in the superfluid. In Table~\ref{t2} we
list the lowest quasiparticle energies for the
solitons in panels (c)-(d) of Fig.\ref{FigDeltas}.
The energies $\omega_{m>1}$ start roughly at the gap
close to the center of the trap $|\Delta(x\approx
0)|$. This is the usual gap in the BCS state.
However, the lowest $\omega_1$ is an order of
magnitude lower. This is a quasiparticle bound state
localized inside the soliton where the gap function
is close to zero. This mode is a soliton counterpart
of the Caroli-de Gennes bound states inside a vortex
core \cite{CdG}. The bound state manifests itself in
the density plots in Fig.\ref{FigRhos} by the fast
density oscillations inside the soliton. The
spectroscopic detection schemes will see the energy
of the bound state in the soliton and they can be
used both to detect the BCS state in 1D and to
distinguish the ground state from the soliton state.

In conclusion, we have explored physics of soliton in
superfluid atomic Fermi gases. Well defined kink-like
soliton in the gap function can be obtained in a
quasi-1D trap. We have estimated its width and energy
and argued that small fluctuations of a soliton
around the center of the trap have positive
frequency. In this respect the BCS soliton is more
robust than the dark soliton in a Bose-Einstein
condensate. The soliton in superfluid fermi gases
can be detected by spectroscopic measurements where
the energy of the quasiparticle bound state localized
in the soliton may be observed.

\begin{table}
\caption{\label{t1} 
Values of $\alpha$ parameter, Eq.~(\ref{alp}), obtained from fitting of a
trial function, $\gamma\Delta_{\rm LDA}(x)\tanh(x/\xi)$, to the gap functions
shown in panels (b)-(d) of Fig.~\ref{FigDeltas}.}
\begin{tabular}{llll}
Panel in Fig.~\ref{FigDeltas} & (b) & (c)  & (d) \\
\hline
$\alpha$ & $1.07$ & $1.05$ & $1.07$    \\
\end{tabular}
\end{table}

\begin{table}
\caption{\label{t2} 
The lowest quasiparticle
energies for the solitons in panels (c)-(d) of Fig.\ref{FigDeltas}.}
\begin{tabular}{lll}
Panel in Fig.~\ref{FigDeltas} & (c)  & (d) \\
\hline
$\omega_1$  &    $1.9$  &    $4.0$   \\
$\omega_2$  &   $11.5$  &   $20.1$   \\
$\omega_3$  &   $11.7$  &   $20.3$   \\
$\omega_4$  &   $13.0$  &   $21.4$   \\
\end{tabular}
\end{table}

{\bf Acknowledgements.---} This work was supported in part by
the the KBN grant PBZ-MIN-008/P03/2003.

\end{document}